# Nanoscale Solid State Batteries Enabled By Thermal Atomic Layer Deposition of a Lithium Polyphosphazene Solid State Electrolyte

*Alexander J. Pearse\*[†], Thomas E. Schmitt[†], Elliot J. Fuller[‡], Farid El-Gabaly[‡], Chuan-Fu Lin[†], Konstantinos Gerasopoulos[⊥], Alexander C. Kozen[⌐], A. Alec Talin[‡], Gary Rubloff[†], Keith E. Gregorcyzck\*[†]*

[†]Department of Materials Science and Engineering, University of Maryland, College Park, MD 20740

[⊥] Research and Exploratory Development Department, The Johns Hopkins University Applied Physics Laboratory, Laurel, MD 20723

[⌐]American Society for Engineering Education, residing at the U.S. Naval Research Laboratory 1818 N St NW, Suite 600 Washington DC, 20036

[‡]Sandia National Laboratories, Livermore, CA 94551

**ABSTRACT:**

Several active areas of research in novel energy storage technologies, including three-dimensional solid state batteries and passivation coatings for reactive battery electrode components, require conformal solid state electrolytes. We describe an atomic layer deposition (ALD) process for a member of the lithium phosphorus oxynitride (LiPON) family, which is employed as a thin film lithium-conducting solid electrolyte. The reaction between lithium *tert*-butoxide (LiO$^t$Bu) and diethyl phosphoramidate (DEPA) produces conformal, ionically conductive thin films with a stoichiometry close to $Li_2PO_2N$ between 250 and 300C. The P/N ratio of the films is always 1, indicative of a particular polymorph of LiPON which closely resembles a polyphosphazene. Films grown at 300C have an ionic conductivity of 6.51 ($\pm$0.36) $\times 10^{-7}$ S/cm at 35C, and are functionally electrochemically stable in the window from 0 to 5.3V vs. Li/Li+. We demonstrate the viability of the ALD-grown electrolyte by integrating it into full solid state batteries, including thin film devices using $LiCoO_2$ as the cathode and Si as the anode operating at up to 1 mA/cm$^2$. The high quality of the ALD growth process allows pinhole-free deposition even on rough crystalline surfaces, and we demonstrate the fabrication and operation of thin film batteries with the thinnest (<100nm) solid state electrolytes yet reported. Finally, we show an additional application of the moderate-temperature ALD process by demonstrating a flexible solid state battery fabricated on a polymer substrate.

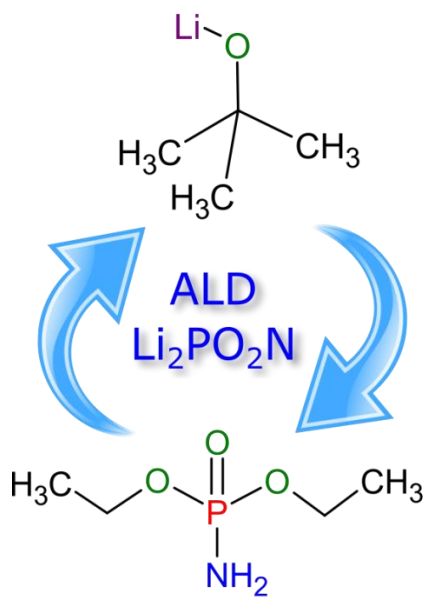
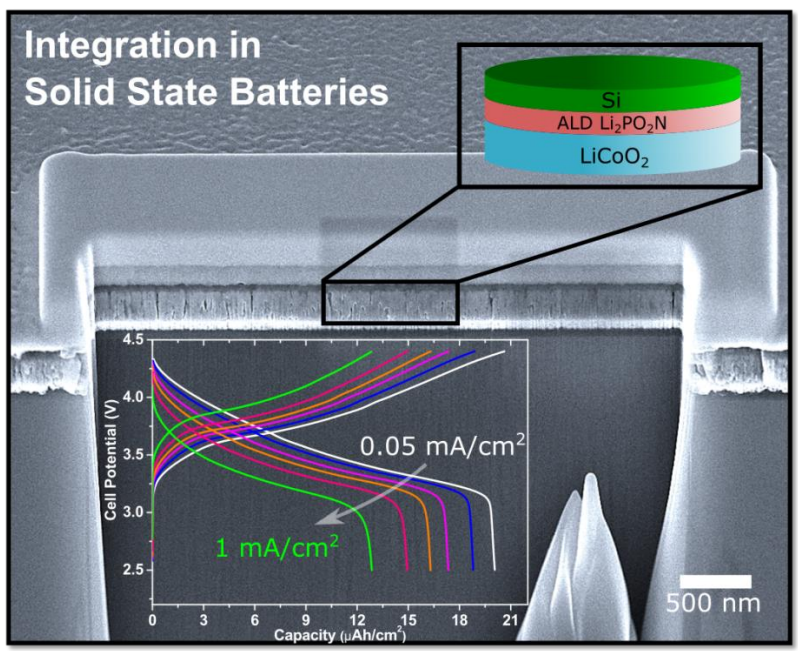

# Introduction

Lithium-ion conducting solid state electrolytes (SSEs) are increasingly important materials in the energy storage technology landscape. SSEs enable all-solid-state secondary lithium-based batteries (SSBs) by directly replacing the flammable organic liquid electrolytes currently used in lithium-ion secondary cells, which significantly reduces or eliminates the chance for catastrophic failure.[1–3] As a result, SSBs are particularly attractive for safety-critical applications, such as aircraft power systems or human-integrated wearable or implantable electronic devices. SSEs also enable, in some cases, wider voltage windows through increased electrochemical stability[4], better high-temperature stability[5], and even increased power density though the higher concentration of charge carriers available in SSEs.[6] SSEs are also playing an increasingly important role as passivation coatings for electrodes and electrode particles,[7–9] both in conventional liquid-based lithium ion systems and in solid state systems. Thin SSEs promote the stable coupling of otherwise reactive cell components, such as Li metal and water,[10] or various cathode materials and sulfide-based solid electrolytes.[6]

The ability to grow SSEs *conformally* is particularly advantageous, i.e. with uniform thickness over challenging three-dimensional (3D) topography. The only currently commercially available SSBs are thin-film solid state batteries, which have many attractive qualities including stability for thousands of cycles, excellent electrode/electrolyte interface quality, and extremely low self-discharge rates.[11,12] However, as thin film SSBs are currently exclusively made using line-of-sight deposition techniques such as thermal evaporation and RF sputtering, their fabrication is limited to planar substrates, ultimately placing an upper limit on their energy density on the order of ~1 mWh/cm$^2$.[11] The ability to grow 3D thin film SSBs on high surface area patterned substrates using conformal deposition processes would alleviate this limitation and allow for the independent design of both power and energy densities per device footprint.[13–15] Attempts to accomplish this with sputtering have been largely unsuccessful due to electrical shorts and inhomogeneous current distributions.[16,17] In the context of passivation coatings, conformally-grown SSEs are required to cover the complex, 3D structure of both individual electrode particles and preformed composite electrodes.[18]

In this work, we report the development of a SSE in the lithium phosphorus oxynitride (LiPON) family grown using atomic layer deposition (ALD), which utilizes self-limiting gas-phase chemical reactions to grow thin films of material.[19] This property enables ALD to grow extremely conformally, to avoid the interaction of supporting solvents with substrates, and often to allow for lower deposition temperatures when compared with chemical vapor deposition. Previously demonstrated ALD electrolytes include Li-containing amorphous metal oxides made by combining a lithium oxide ALD process with existing multicomponent oxide processes, including Li-Al-O, Li-Al-Si-O, Li-La-Ti-O, Li-Nb-O, and Li-Ta-O ternary and quaternary films.[20–24] These processes uniformly produced materials with ionic conductivities of < $10^{-7}$ S/cm at room temperature, grow slowly due to the number of ALD subcycles involved, and often incorporated multivalent metal ions which can degrade electrochemical stability. Finally, a sulfide electrolyte ($Li_xAl_yS_z$) has also been demonstrated.[25]

The most promising ALD electrolytes to emerge to date are members of the LiPON family, which is the electrolyte of choice in existing thin film batteries due to its electrochemical stability, ionic conductivity (~$10^{-6}$ S/cm) and high electrical resistivity.[12,26,27] Incorporating nitrogen into existing ALD processes[28] for $Li_3PO_4$ proved to be a challenge. The first ALD process for LiPON involved[29] nitrogen incorporation through use of a $N_2$ plasma, which, while providing an attractive degree of compositional tunability, induces limits on conformality due to plasma radical recombination in high aspect ratio structures. Nisula et al. introduced, nearly simultaneously, the use of diethyl phosphoramidate (DEPA) as a precursor,[30] which contains a pre-formed P-N bond, and grew LiPON-family films with a stoichiometry of $Li_{0.9}P_1O_{2.8}N_{0.55}$ with some hydrocarbon incorporation at 290C using lithium hexamethyldisilazide (LiHMDS) as a lithium source. Shibata also recently demonstrated[31] a thermal process for LiPON using $NH_3$ as a nitrogen source along with lithium *tert*-butoxide (LiO$^t$Bu) and tris-dimethylaminophosphorus, but reported growth only at temperatures well above the thermal decomposition temperature of LiO$^t$Bu, which calls into question the self-limiting nature of the process.[32] Plasma-enhanced chemical vapor deposition processes for LiPON-family films have also been developed.[33] In this report, we explored the reaction between LiO$^t$Bu and DEPA, which results in the growth of conformal, high quality solid electrolytes with a stoichiometry of ~ $Li_2PO_2N$ (excluding carbon contamination) which we identify as a lithium polyphosphazene (LPZ) for reasons discussed below.

Despite several reports of ALD-based SSEs, none have been tested when integrated into full solid state batteries, nor have the electrochemical stability windows been established in most cases. While metal-electrolyte-metal stacks allow for the characterization of ionic and electronic conductivity, they do not simulate realistic electrode/electrolyte interfaces, which are often chemically and electrochemically reactive, and quite rough in the case of crystalline electrodes. A major benefit of very thin SSEs is that their total resistance can be low enough such that a battery will be limited in power performance by ionic diffusion in the anode/cathode well before Ohmic losses in the electrotye.[17,34] Achieving higher power performance then requires increasing the surface area of the electrodes using 3D architectures, for which a conformal solid state electrolyte is necessarily required. In addition, when using ultrathin solid electrolytes, the electrochemical stability is arguably more important than the ionic conductivity. In contrast to sputtered materials, ALD- grown electrolytes often contain fragments of precursor ligands from incomplete reactions, and their effect on electrochemical stability is unclear. [21,30] Finally, while ALD is generally considered capable of growing electronically insulating films at lower thicknesses than any other film deposition technique, the downscaling capability of ALD-grown solid electrolytes in complete batteries is untested. Here, we show that ALD LPZ is compatible with two different solid state battery chemistries ($LiCoO_2$/Si and $LiV_2O_5$/Si), fully characterize its transport characteristics and electrochemical stability, and demonstrate the thinnest reported solid electrolyte (~35nm) in a realistic full battery (>3V cell voltage) holding charge.

## Results and Discussion

### The Lithium *tert*-Butoxide and DEPA ALD Process

**Process Parameters:** We explored the growth characteristics and chemistry of the LiO$^t$Bu-DEPA reaction between 200 and 300C primarily using two *in-situ* methods. First, we utilized *in-situ* spectroscopic ellipsometry (SE) to noninvasively determine the process parameters, and growth rate of the deposited thin films. Second, we utilized x-ray photoelectron spectroscopy (XPS) to determine the detailed chemistry of the deposited films. The ALD reactor and XPS system are coupled through an ultrahigh vacuum transfer chamber (Figure S1), allowing for the rigorous exclusion of surface contamination. As the grown polyphosphazene films were found to be sensitive to air exposure, these two techniques provide the most reliable information.

SE measures changes in the polarization of light upon reflection from an optically flat surface, and when an appropriate optical model is determined, can easily measure sub-monolayer thickness changes as they occur pulse-to-pulse during an ALD process.[35] For ALD development, in-situ SE has the further advantage of rapid process characterization, as it is possible to vary parameters such as temperature and pulse times while monitoring deposition on a single substrate and without breaking vacuum. Details of the optical model used here can be found in the experimental section.

Overall, we find that the LiO$^t$Bu-DEPA reaction is a well-behaved, though non-ideal, ALD process, exhibiting self-limiting growth as a function of precursor dosage but lacking an obvious temperature window of constant growth rate. Both precursors are solids at room temperature, though we found that only LiOtBu needed the assistance of a bubbler for delivery. Figure 1 outlines the processes parameters as determined by SE. Figure 1a shows a snapshot of typical linear growth measured in-situ for the baseline 300C process. The inset shows differential increases in film thickness associated with both the LiO$^t$Bu and DEPA pulses, resulting in a net growth rate of approximately 0.9 Å/cyc. Figure 1b shows growth rates measured as averages over 30 cycles for different combinations of LiOtBu and DEPA pulse times determined by SE on a single sample after steady state growth had been achieved. Both precursors exhibit self-limiting behavior, indicative of the surface-mediated half-reactions typical of an ALD process. LiO$^t$Bu requires exceptionally long pulse times to saturate, which is due to both its intrinsically low vapor pressure and our inclusion of low-conductance particle filters in the delivery lines to prevent fine particles of precursor from reaching the chamber. However, as shown in Figure 1c, the growth rate increases approximately linearly across the entire tested temperature range, from about 0.15 Å/cyc at 200C to 0.9 Å/cyc at 300C and does not exhibit a constant-growth window, consistent with a thermally activated reaction. A variety of ALD processes are self-limiting but lack a clear constant-growth window.[36]

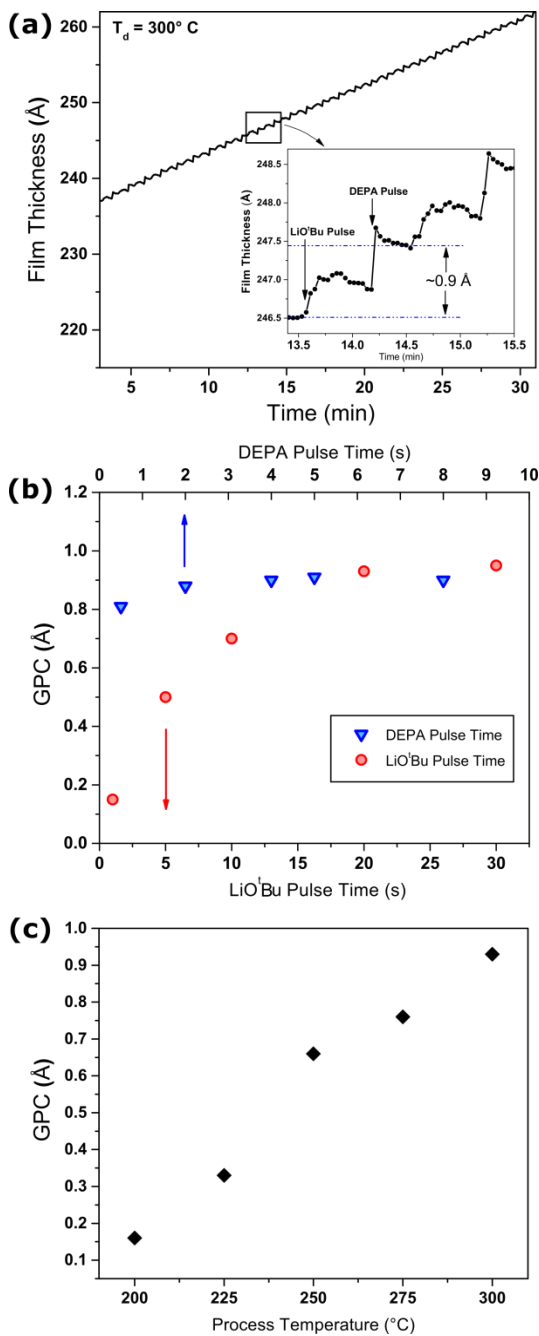

**Figure 1:** Process parameters of the LiOtBu-DEPA ALD reaction measured by in-situ spectroscopic ellipsometry. (a) A snapshot of linear growth at 300C with the inset showing 2 full cycles. (b) Growth per cycle of films at 300C as a function of precursor dose time, showing saturation for both precursors. The LiOtBu pulse time was fixed at 20s and the DEPA pulse time was fixed at 2s when varying the other precursor. (c) Growth rate as a function of reactor temperature

Repeated measurements of the ALD process at 300C has also shown that the growth rate tends to slowly decline over time if the precursors are kept continuously heated. After several weeks of storage on the ALD system, the overall growth rate is often reduced by 20 to 30%, although all other aspects of the ALD process, including self-limiting behavior, are preserved. As the nominal growth rate is restored by replacing the LiO$^t$Bu, we believe that the precursor undergoes a slow decomposition reaction even at moderate temperatures (100 – 140C). Saulys et al. have suggested[32] that LiO$^t$Bu may undergo a self-catalyzed decomposition reaction induced by trace H$_2$O, producing *tert*-butanol and isobutylene, which could contribute to the reduced growth rate if these species adsorb on the substrate surface and block reactive sites.

**Film Characterization:** Films deposited at 250 and 300 C (LPZ-250 and LPZ-300, respectively) were transferred under UHV directly from the ALD chamber to a coupled XPS spectrometer to identify the chemical composition. The spectra, along with the proposed molecular structure of the material, are shown in Figure 2. While measuring Li-containing thin film composition by XPS quantification is normally challenging due to the tendency of many such materials to react with air and other environmental contaminants, forming a compositional gradient within the XPS analysis region, our experimental conditions preserve the surface region and allows for accurate analysis.[37] Table 1 summarizes the composition found through quantification of the high resolution peaks in Figure 2. The films are composed entirely of Li, P, O, N, and C, (Figure 2l) indicating that the ALD process produces a member of the lithium phosphorus oxynitride (LiPON) family. The LiPON family comprises a wide range[38,39] of compositions and microstructures which lie inside a quaternary phase diagram with the endpoints Li$_2$O, Li$_3$N, P$_3$N$_5$, and P$_2$O$_5$. The exact chemistry of a given LiPON-family material has a strong effect on its ionic conductivity, electrochemical stability, and environmental stability.[26,39] The nature of N incorporation is particularly influential, with higher nitrogen concentrations generally correlating with higher ionic conductivity and a lower activation energy.[40,41] When substituted for oxygen in lithium phosphate, nitrogen atoms can link either two (=N-) or three (>N-) phosphorus centers, as identified by the doublet commonly observed in N 1s XPS spectra. No clear correlation has been identified in the literature as to whether one type of bonding is preferable; the analysis is complicated by the fact that most XPS studies are *ex-situ*, and the surface chemistry may not reflect the bulk.

Table 1: XPS quantification of ALD LPZ films. Samples were transferred to the spectrometer under vacuum. Numerical values are atomic percent composition.

|  | Li | P | O | N | C | Comp. Relative to P excluding C |
|---|---|---|---|---|---|---|
| LPZ-250 | 24.1 | 14.3 | 29.9 | 14.9 | 16.8 | Li$_{1.7}$P$_1$O$_{2.1}$N$_1$ |
| LPZ-300 | 27.9 | 15.0 | 32.0 | 15.2 | 10.0 | Li$_{1.9}$P$_1$O$_{2.1}$N$_1$ |

XPS quantification (Table 1) shows that the ALD process produces a composition close to the stoichiometry Li$_2$PO$_2$N, especially at higher temperatures, albeit with a significant amount of carbon incorporation from residual ligands. In particular, the P/N atomic ratio in these films is always 1 to within the accuracy of XPS quantification. The stoichiometry Li$_2$PO$_2$N strongly suggests a particular polymorph of LiPON, recently predicted to be stable and synthesized by Du et al.[42] in a crystalline form, in which alternating P and N atoms form a linear backbone with Li atoms coordinating with both oxygen atoms bonded to the phosphorous centers and the linking nitrogen atoms (Figure 1a).

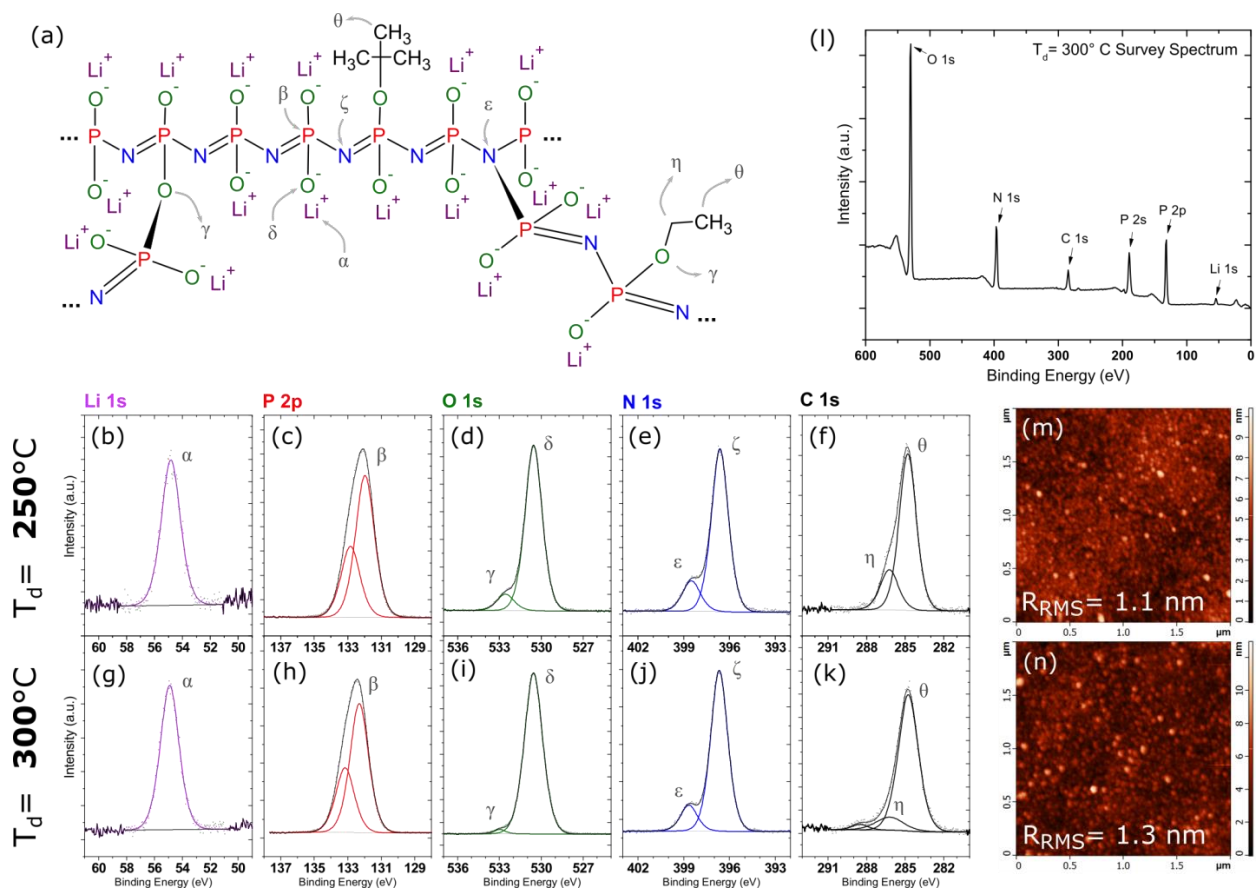

**Figure 2:** (a) A schematic of the proposed molecular structure of the ALD-grown Li$_2$PO$_2$N (ALD LPZ). Individual atomic sites are labelled with greek letters which correspond to peaks identified in the high resolution XPS data shown below the schematic. (b-k) High resolution XPS core level spectra of ALD LPZ films grown at 250 and 300C. Spectra are calibrated to the θ component of the C 1s at 284.8 eV in each case. (l) Survey spectrum of a LPZ film grown at 300C showing the relative intensities of the constitutive elements (m,n) Tapping mode AFM of ~50nm ALD LPZ films grown at 250 and 300C showing a root mean square roughness (R$_{RMS}$) of below 2nm in each case.

Stoichiometries approaching Li$_2$PO$_2$N have been achieved with RF sputtering processes in a few reports, generally when performed in pure N$_2$.[39,41] Crystalline Li$_2$PO$_2$N has also been used as a sputtering target for depositing electrolyte thin films.[43] Because of the presence of linear P-N chains, we refer to this polymorph as a "lithium polyphosphazene" (LPZ) to differentiate it from the broader term "LiPON". Li$_2$PO$_2$N sits at the boundary between conventional LiPON glasses and polyphosphazene-based salt-in-polymer electrolytes, which have been explored for their chemical stability in lithium-ion batteries.[44,45] The films are always amorphous in the tested temperature range, as indicated by the lack of identifiable peaks in XRD (Figure S7) and the few-nm surface roughness as measured by AFM (Figure 2m, 2n).

The high resolution XPS spectra also strongly support the identification of the grown material as a lithium polyphosphazene containing a population of chemical defects. Figures 2b-2k show core level spectra from films grown at two different temperatures, in which fitted chemical components are labelled with greek letters corresponding to proposed associated atomic sites labelled in Figure 2a. These data were repeatable and consistent for a given deposition temperature. All core level spectra are calibrated by placing the lower binding energy component of the C 1s spectra at 284.8 eV under the assumption that this peak is associated with embedded hydrocarbon from residual ligands and unreacted precursor fragments. In the ideal polyphosphazene chain structure, there is only one distinguishable chemical environment for each of Li (-O$^-$Li$^+$), P (=P-), O (P-O$^-$Li$^+$), and N (-N=). In the XPS data, there is indeed only one component identifiable in the Li 1s and P 2p spectra, designated as the α and β components (note that the P 2p spectra are fit with a constrained spin-orbit split doublet). The O 1s and N 1s spectra contain two components, each with a minor impurity peak on the high binding energy side of a major component. For each, we identify the larger O 1s δ and N 1s ζ components as originating from the primary polyphosphazene chain structure, and the much smaller γ and ε components as originating from a number of possible chemical defects. Theoretical calculations by Du et al. and far-IR spectroscopy by Carrillo Solano et al. have indicated that Li cations in the $Li_2PO_2N$ structure coordinate with both the O and N atoms, likely creating a weak or partial ionic bond with both.[38,39] This is consistent with the relatively low binding energy of the O 1s δ at 530.6 eV and especially the N 1s ζ component at 396.7 eV, which sits in a range normally associated with $N^{3-}$ in metal nitrides.[46] These binding energies are generally in agreement (within 1 eV) with other XPS measurements[40] of LiPON, though the comparability of data taken from air exposed films is questionable given that virtually all forms of LiPON are air reactive through hydrolysis and carbonate formation.[47]

Next, we identify the origin of the O 1s γ and N 1s ε components, as well as the nature of the carbon incorporation. In the LiPON family, the N 1s peak is commonly split into two components, with a lower binding energy peak associated with doubly bonded N (P-N=P) and a high binding energy peak associated with triply bonded N (P-N$<^P_P$), in general agreement with the spectral shapes observed in Figures 2e and 2j and with the typical 1.5 eV separation between the ε and ζ components.[39] The N 1s ε component is therefore tentatively identified as triply bonded nitrogen, forming links between linear polyphosphazene chains, and decreases in intensity for LPZ-300 relative to LPZ-250. The O 1s γ component, located at about 532.6 eV for LPZ-250 and 533 eV for LPZ-300, lies in a crowded region of binding energies which includes many organic oxygen-carbon species as well as phosphorus-bridging oxygen (P-O-P), which is commonly observed in amorphous phosphates and LiPON with a N/P ratio of less than 1.[29,40] This peak is most likely linked to the C 1s η component which sits between 286 and 287 eV, consistent with -C-O- bonding. Taken together, the O 1s γ and C 1s η and θ components all primarily arise from precursor ligands incorporated into the film, including *tert*-butoxy (-OC(CH$_3$)$_3$) groups from the Li precursor and diethyl groups from DEPA. We also note that the primary deviation from the ideal stoichiometry ($Li_2PO_2N$) is a Li deficiency, especially for films grown at 250C. This can be rationalized by recognizing that incorporated organic ligands from the precursors would likely replace the –OLi group on the phosphazene chain, leading to an inverse correlation between Li and C content, as is observed. Finally, the carbon content and impurity components all decrease in relative intensity with the higher growth temperature, indicating a more phase-pure material with longer polyphosphazene chains on average.

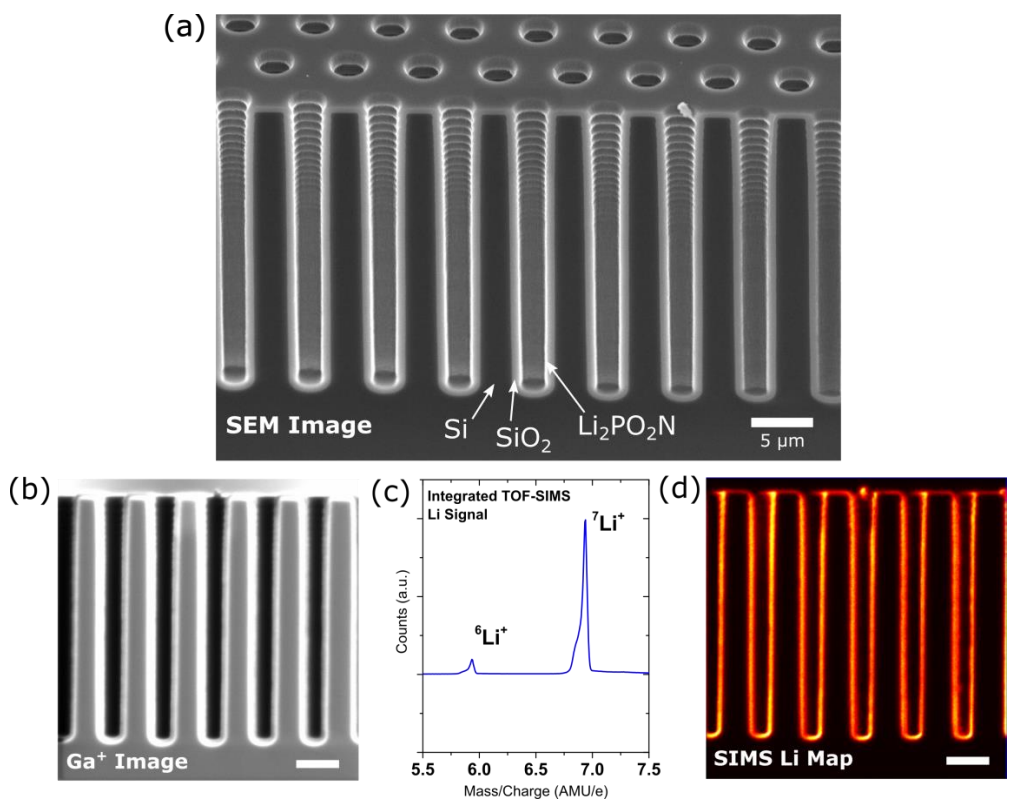

**Figure 3:** (a) SEM image of RIE etched holes (aspect ratio 10) in Si on the side of a cleaved wafer. (b) Ga+ (FIB) image of a targeted region for analysis (c) Li signal from FIB-excited TOF-SIMS which is then mapped in (d), showing the distribution of the Li-containing ALD LPZ film down the hole walls. Small variations in signal intensity are primarily a result of geometric effects relating to the orientation of the FIB, the sample, and the SIMS detector. Scale bars in all images correspond to 5 μm. No Li signal is observed at the bottom of the holes due to a shadowing effect from the 3D geometry.

The production of $Li_2PO_2N$ suggests that the ALD reaction between LiO$^t$Bu and DEPA is complex, and we do not propose a complete mechanism at this time. The atomic ratio N/P = 1 in the product suggests that the P-N bond in the DEPA molecule is not broken during the ALD reactions, and the ratio O/P = 2 in the product compared to O/P = 3 in DEPA indicates oxygen is lost through the breaking of a P-O bond, which is surprising given that P-N bonds are generally considered to be weaker and more reactive.[48] The ALD reaction reported here also appears to be chemically distinct from the LiHMDS-DEPA ALD process reported[30] by Nisula at al., which produced a significantly different LiPON-family thin film with the stoichiometry $Li_{0.9}P_1O_{2.8}N_{0.55}$ at 290C, more closely resembling a nitrogen-substituted lithium metaphosphate ($LiPO_3$) than a polyphosphazene when the Li/P and N/P ratios are compared. This compositional difference, along with entirely different temperature-dependent growth rates of the LiHMDS-DEPA reaction, suggests that the ligand chemistry of the Li precursor plays a significant role in the ALD reaction pathways. If the structure of the ALD-grown films is indeed polyphosphazene chains, we believe that these reactions may be better characterized as surface-mediated polymerizations rather than a traditional ligand-exchange ALD process, as the DEPA molecules must be linked through the amine group to achieve the ending stoichiometry. Nielsen demonstrated head-to-tail self-condensation of diphenyl phosphoramidate forming P-N-P chains in the presence of a strong base which could deprotonate the amine group, and we believe the LPZ ALD reaction could follow a similar pathway.[49] This linking process is not purely thermally activated and must involve LiO$^t$Bu, as in-situ SE of repeated pulses of DEPA alone at 300C shows no significant film growth (Figure S2). Atanasov et al. reported oxidative polymerization in a molecular layer deposition process, indicating the possibility of such surface-controlled polymerization reactions. [50]

**Conformality**

An important feature of ALD is the ability to conformally coat high aspect ratio structures. To demonstrate the conformality of the LiOtBu-DEPA ALD process, we fabricated arrays of holes, 3 microns in diameter and 30 microns deep, etched into a Si wafer using reactive ion etching (Figure 3a). The hole array was thermally oxidized to form a $SiO_2$ Li diffusion barrier and exposed to 910 cycles of the LiOtBu-DEPA ALD process at 300C. We found that the deposited polyphosphazene films were difficult to clearly image on a cleaved cross-sectional sample and showed very little contrast with $SiO_2$, and therefore measured the spatial Li distribution directly using $Ga^+$ excited time-of-flight secondary ion mass spectroscopy (ToF-SIMS) (Figure 3b-d). The ToF-SIMS analysis clearly shows the presence of a Li-containing thin film all the way to the pore bottom, demonstrating conformality in a structure with an aspect ratio of 10.

**Transport Characteristics and Stability of ALD LPZ**

To test the ALD LPZ films for suitability as SSEs in thin film planar and future 3D SSBs, we measured the ionic and electronic conductivities of the material in several configurations (Figure 4). First, we fabricated metal-electrolyte-metal (MEM) stacks using electron-beam evaporated Pt as a symmetric blocking electrode (Pt/$Li_2PO_2N$/Pt), using a planar Pt film as the bottom electrode and shadow-masked 1mm diameter circular top electrodes to define the device area. Potentiostatic electrochemical impedance spectroscopy (PEIS) was used to measure the ionic conductivity of LPZ films grown at 300C and 250C (80nm and 70nm in thickness, respectively, measured by FIB cross section). The LPZ films were air-exposed for approximately 10 minutes during fabrication before anode deposition. Figure 4a shows results from the impedance tests plotted on the complex plane for the LPZ-300 film at three different temperatures, as well as data from the LPZ-250 sample at 35C. While all the spectra exhibit the hallmarks of ionic conductivity, which include semicircular arcs at high frequencies followed by a rapid increase in the imaginary component of the impedance at low frequencies concurrent with double layer formation on the blocking electrodes, the data also indicate two separate components in the high frequency region. This suggests two separate ionic transport processes, possibly due to to either grain boundary transport or the presence of a reactive interphase layer at the electrode-electrolyte interface. As the LPZ is amorphous, we adopted the second explanation and modelled this data using the equivalent circuit shown in Figure 4a. The model includes two parallel resistor/constant phase element (CPE) components in series, with one ($R_b$ and $CPE_b$) corresponding to the "bulk" of the LPZ film and the second ($R_r$ and $CPE_r$) corresponding to the resistive reaction layer. A third CPE ($CPE_W$) models the Warburg-like blocking behavior at low frequencies. We found the use of constant phase elements, which empirically take into account the distribution of activation energies and correlated ion motion expected from an ionically conductive glass,[51] necessary to fit the data. Detailed fitting parameters for the model can be found in Table S1. The total ionic conductivity, calculated from $\sigma^{-1} = (R_r + R_b)A/L$, where $A$ is the electrode area and $L$ is the film thickness, is $1.6 (\pm 0.1) \times 10^{-7}$ S/cm for LPZ-300 and $1.4 (\pm 0.14) \times 10^{-7}$ S/cm for LPZ-250, measured at 35C, indicating that the ionic conductivity decreased slightly at the lower deposition temperature, consistent with the larger amount of impurities identified by XPS. This value is lower than typical sputtered LiPON by roughly one order of magnitude.[11,26]

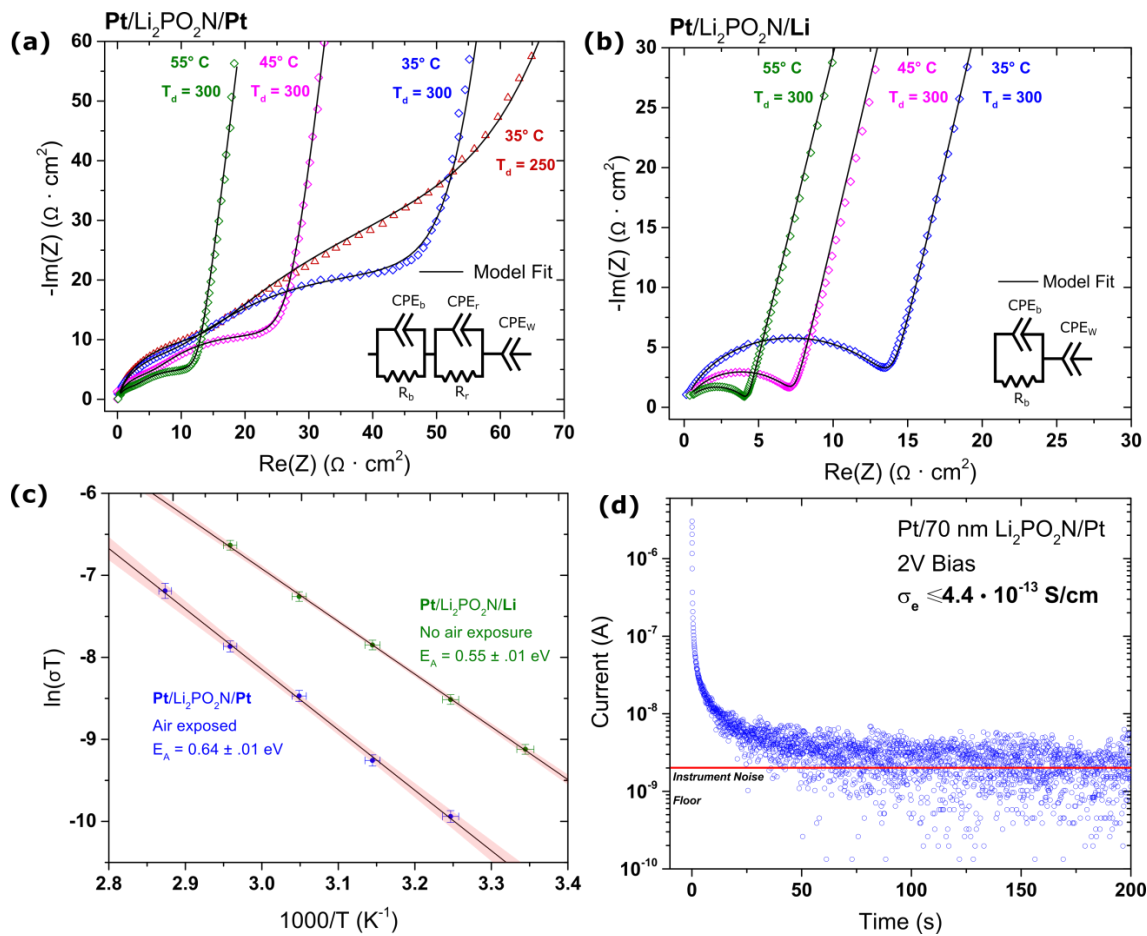

**Figure 4:** Transport measurements of ALD LPZ. (a) PEIS of Pt/80nm LPZ-300/Pt and Pt/70nm LPZ-250/Pt film stacks. The data for LPZ-300 ($T_d$ = 300) is plotted at several temperatures to illustrate the thermally activated transport process. For these samples, the LPZ was briefly air-exposed, leading to the development of a second arc at medium frequencies. (b) PEIS of Pt/LPZ-300/Li synthesized without air exposure, demonstrating an ideal single arc at high frequencies and an overall higher conductivity. (c) Activation energies for ionic transport in LPZ-300 films, with and without air exposure. (d) Current-time response from a 2V constant bias in a Pt/70nm LPZ-250/Pt stack. LPZ-300 films showed a similar response. The red line shows the lower limit of resolution for the potentiostat. No dielectric breakdown or increase in conduction was observed even after one hour of polarization.

XPS depth profiling of an air-exposed LPZ film using an argon gas cluster sputtering source revealed the presence of a significant layer of $Li_2CO_3$ formed at the LPZ/air interface, as well as changes in the N chemistry near the surface (Figure S3). $R_r$ is identified as originating from this surface region because of the lower ionic conductivity of $Li_2CO_3$.[52] Similar chemical reactions have been identified in LiPON-family materials (as well as many other Li solid electrolytes) previously.[30,47] While LiPON stoichiometries near $Li_2PO_2N$ have been previously reported[39,42] to be remarkably air-stable, we unfortunately find that the ALD-grown LPZ does not maintain this stability, possibly as a result of reactions involving the incorporated organic ligands and the amorphous, more chemically defective structure.

We also fabricated Pt/$Li_2PO_2N$/Li solid state half cells using LPZ-300 to test the electrochemical stability of the deposited films. As the lithium evaporator used is directly coupled to the vacuum system used for ALD growth, we were able to make the full stack without any environmental exposure. PEIS in this case exhibited a much more ideal response (Figure 4b) and could be fit with only a single R/CPE component, which further confirmed that the extra impedance $R_r$ was due to a decomposition layer in the Pt/$Li_2PO_2N$/Pt devices and allowed us to isolate the true bulk resistivity $R_b$. The ionic conductivity was found to be $6.51\,(\pm 0.36) \times 10^{-7}$ S/cm when measured at 35C, which is comparable to that found by Nisula et al. for the LiHMDS-DEPA ALD process despite the difference in composition, and is among the highest values measured for ALD grown solid electrolytes.[25,29,30] All devices tested exhibited decreasing impedance with increasing temperature (Figure 4a,4b), and we measured the activation energy $E_A$ from the Arrhenius relation $\sigma T = A\exp[-\frac{E_A}{kT}]$, plotted in Figure 4c. Air-exposed LPZ-300 films had an effective (including effects from the reaction layer) activation energy of $0.64 \pm 0.01$ eV, in contrast with the non-air-exposed LPZ-300 devices with $E_A = 0.55 \pm 0.01$ eV. The latter value is in excellent agreement with a number of previous studies of LiPON-family materials.[26,41] Finally, we measured the electronic conductivity of the films by applying a constant 2V bias to Pt/Li$_2$PO$_2$N/Pt stacks (Figure 4d). The measured current rapidly relaxes to the limit of detection of the potentiostat (1 nA), placing an upper bound on the electronic conductivity of $\sigma_e \leq 4.4 \times 10^{-13}$ S/cm, nearly 6 orders of magnitude below the ionic conductivity.

The electrochemical stability of ALD-grown solid electrolytes remains largely unknown. The inclusion of impurities such as reactive defects and residual ligands could degrade electrochemical stability, which is particularly important for very thin SSEs.[16] Figure 5 shows cyclic voltammetry of a Pt/Li$_2$PO$_2$N/Li stack made using 90nm of LPZ-300 where Pt is used as the working electrode against the Li counter and reference. The large negative current below approximately 0.1 V vs. Li/Li+ is associated with Li-Pt alloy formation, and the two peaks appearing on the positive scan involve the subsequent dealloying of Li.[53,54] However, the first cycle Coulombic efficiency of this reaction is only 82%, suggesting some cathodic decomposition of the Li$_2$PO$_2$N simultaneously occurred. No features indicating oxidative decomposition can be observed at this scale. We cycled a separate sample in the range 1-6V vs. Li/Li+ (inset) to explore the oxidative stability in more detail, revealing a small anodic peak at 3.8V followed by what is likely a minor ($\approx$1 µAh/cm$^2$ in total) decomposition reaction. The portion of the decomposition reaction below approximately 5.3V vs. Li/Li+ does not recur after the first cycle (Figure 5, inset). Taking these results together, there are minor decomposition reactions below 0.1V and above 3.8V, consistent with recent theoretical calculations on the stability of LiPON-family materials, indicating a nominal stability window of 0.1-3.8V vs. Li[55] However, LPZ-300 remains functionally ionically conductive and electrically insulating in the whole range of 0-5.3V vs. Li/Li+. These reactions seem thus to form a self-limiting "solid electrolyte-electrode interphase", similar to analogous reactions between organic liquid electrolytes and lithium-ion electrodes, which enables operation of the solid electrolyte in a voltage window wider than its intrinsic thermodynamic stability.

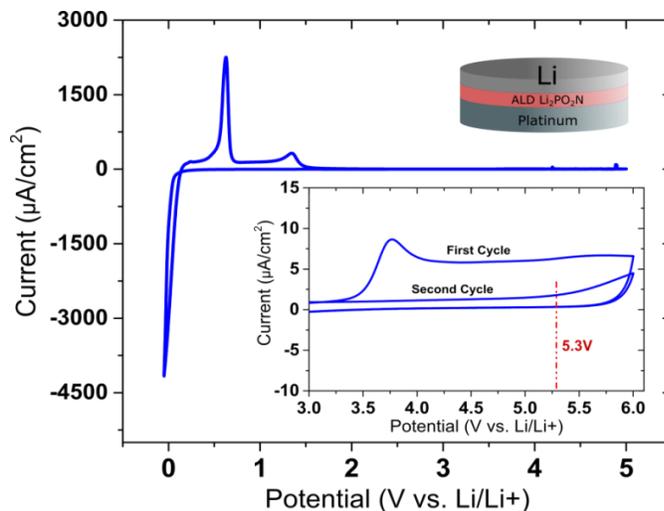

**Figure 5:** Cyclic voltammetry between -0.05 and 5V vs. Li/Li+ of a Pt/90nm LPZ-300/2600nm Li solid state half cell. The sweep rate was 5 mV/s. The inset shows a view of activity during two cycles in the high potential region of a separate sample cycled between 1 and 6V vs. Li/Li+ at the same rate. Note the large difference in y-scale.

**Testing in Solid State Batteries**

The ionic conductivity, electronic resistivity, conformality, and electrochemical stability indicate ALD-grown $Li_2PO_2N$ is an excellent candidate for use in thin film solid state batteries (TSSBs). We fabricated Li-ion solid state full cells using sputtered crystalline $LiCoO_2$ (LCO) as the cathode, ALD LPZ-300 as the electrolyte, and electron-beam evaporated Si as the anode. Pt and Cu thin films were used as cathode and anode current collectors, respectively. Si is a promising replacement for Li anodes, as it does not melt at typical solder-reflow temperatures when used in on-chip energy storage systems and is easier to process and handle.[56,57] In addition, as all-ALD 3D TSSBs will likely not be able to use Li as an anode for lack of a plausible Li ALD process, constructing planar Li-ion TSSBs is a more representative test of the performance of ALD $Li_2PO_2N$ in one of its most promising future applications. Figure 6 shows transmission electron microscopy (TEM) images of a FIB-fabricated cross section of a 500nm LCO/90nm LPZ-300/ 80nm Si battery stack. Of note is the columnar structure of the LCO and the intimate, conformal contact between the ALD LPZ and LCO despite the large interface roughness. Because we observed no visual contrast between LPZ and Si using TEM, we used EELS mapping of O, N and Co to confirm thicknesses of the deposited layers (Figure 6b). The Si anode contains a small amount of $SiO_x$, incorporated during evaporation, which is the origin of the O K-edge intensity from this region. The LPZ eventually decomposes under electron beam exposure, as found in other TEM studies[58] of LiPON, which also allows for easy visual identification of the layer (Figure S4).

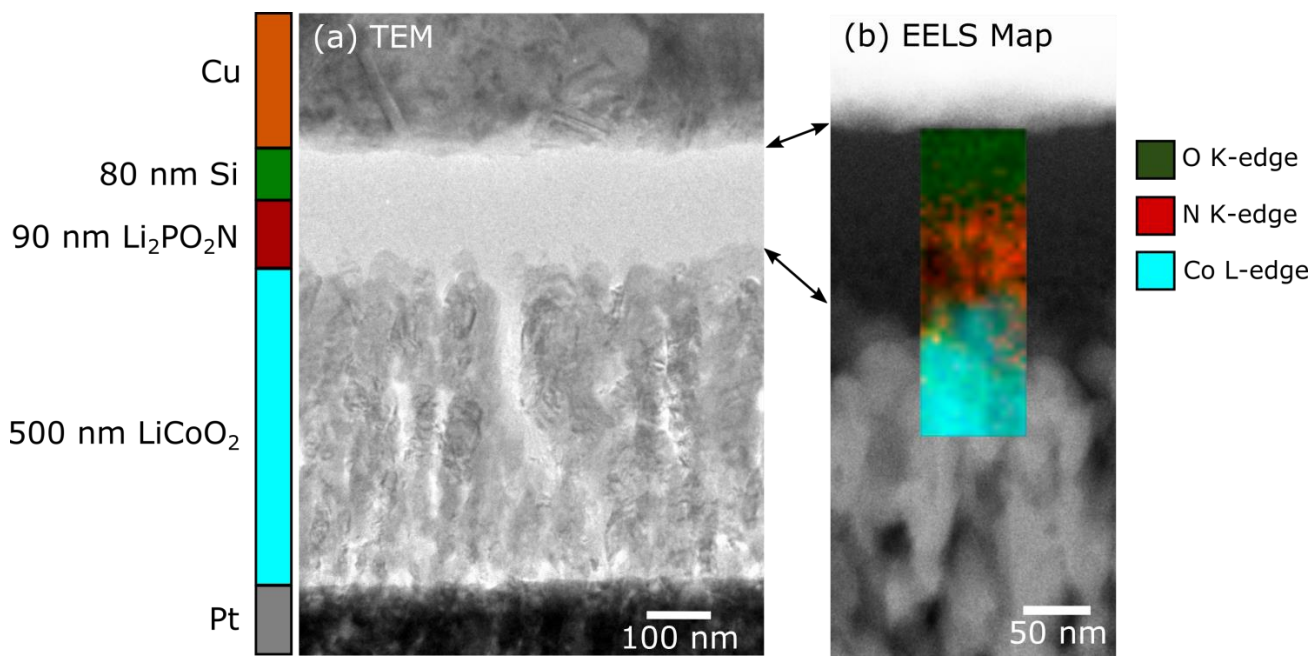

**Figure 6:** TEM characterization of a Pt/LCO/LPZ-300/Si/Cu solid state battery. (a) TEM of a FIB-cut cross section showing the crystalline microstructure of the LCO and an amorphous region containing both LPZ-300 and Si, which cannot be resolved by TEM alone. (b) An EELS map of the elemental distribution of O, N, and Co, which discriminates the LPZ-300 and Si in the amorphous region. The Si anode contains O from both contamination of the Si during evaporation as well as oxidation of the TEM lamella surface.

The electrochemical performance of the battery characterized in Figure 6 is shown in Figures 7a-7c. Three cycles of cyclic voltammetry between 2 and 4.4V, shown in Figure 8a, illustrate the general electrochemical properties of the LCO/Si couple. The peak shapes reflect a convolution of the electrochemical response of both LCO and Si, which leads to broadening and larger anodic/cathodic peak separation when compared to similar data for LCO/Li TSSBs.[53,56] The labelled peaks characteristic of LCO include (1) the two-phase hexagonal I to hexagonal II transition, (2) the two-phase hexagonal II to monoclinic transition, (3) a monoclinic to hexagonal transition.[53,58] Peak (4) is only present during the first cycle, and we believe this represents an irreversible reaction occurring at the LCO-LPZ interface as the location (4.1V) is roughly consistent with the minor decomposition reaction identified in Figure 5. This reaction does not appear to significantly impede subsequent battery operation. Peak (1) shifts to a higher potential and drops in magnitude after the first cycle. We attribute this to an irreversible reaction in the Si anode, associated with both the conversion reaction of contaminant $SiO_x$, which occurs at a higher potential vs. Li/Li$^+$ than Si/Li alloying, as well as Li trapping in the Si itself.[17]

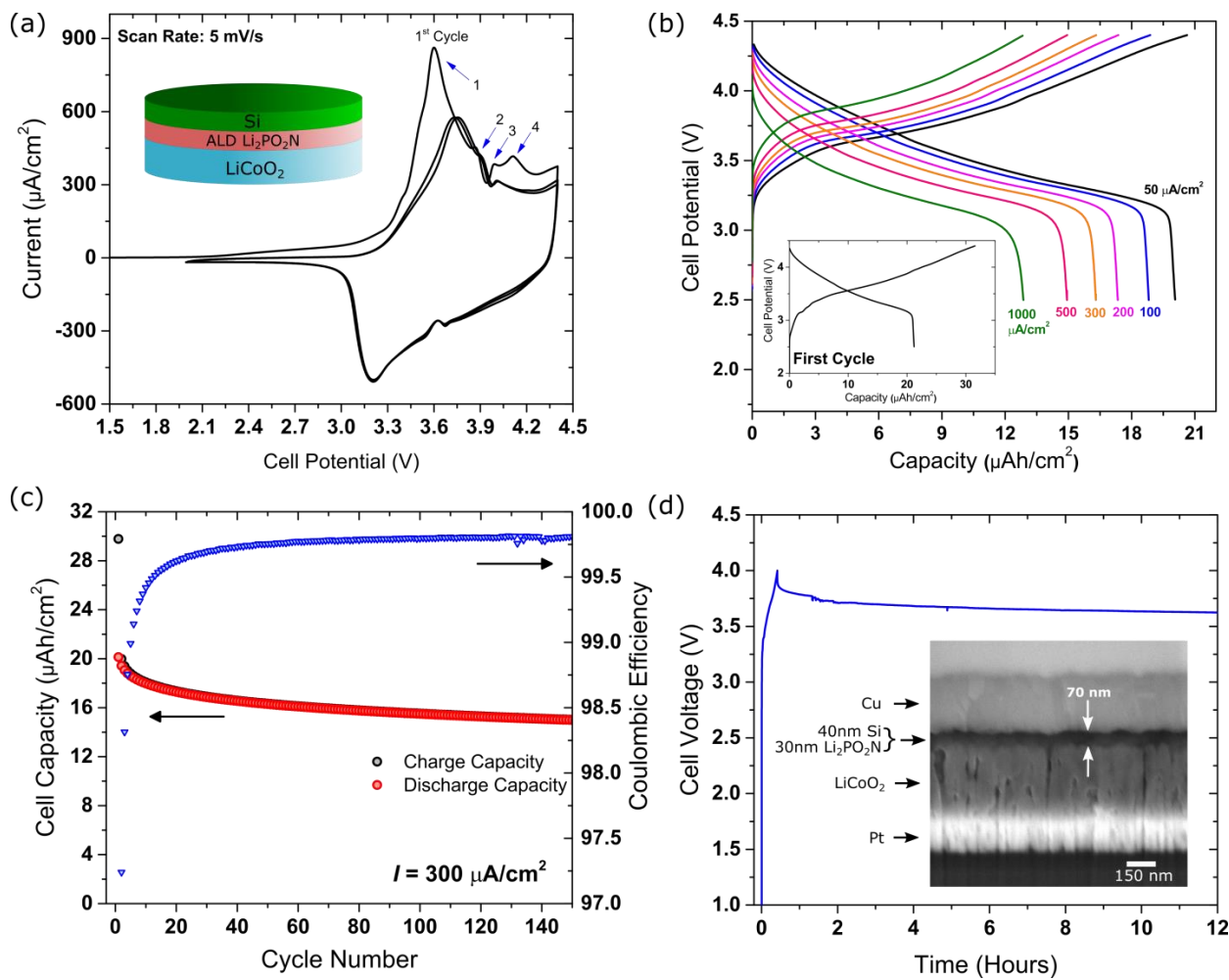

**Figure 7:** Electrochemical characterization of LCO/Si based batteries. (a) Cyclic voltammetry at 5 mV/s of a 500nm LCO/ 90nm LPZ-300/80nm Si battery stack between 2 and 4.4 V showing the first three cycles. Labelled peaks 1, 2, and 3 are associated with delithiation of the LCO, while peak 4 is possibly associated with electrolyte decomposition. (b) Rate performance of the battery measured in (a), with the first cycle irreversible capacity shown in the inset. (c) Cycling stability of a cell cycled at 300μA/cm$^2$ (8.6C based on the theoretical LCO capacity). The capacity stabilizes at approximately 50% of the initial reversible capacity. (d) Charge retention over 12 hours of a sample with a 400 cycle LPZ-300 solid electrolyte, measured to be approximately 30-35 nm in thickness by FIB-SEM cross section. The battery was charged to 4V at a 2C rate, and the OCV was subsequently measured. Samples with ultrathin <40nm LPZ (as shown in (d)) often exhibited dielectric breakdown at cell potentials above 4V, but were stable below this value.

The low total resistance of thin films of LPZ (~56 Ω·cm$^2$ for 90 nm of air-exposed LPZ-300 in this case) allows for excellent rate performance, as shown by the galvanostatic charge/discharge curves in Figure 7b. Although the first cycle (inset) typically engenders a ~33% irreversible capacity loss associated with the Si anode, the battery is able to subsequently deliver 20 μAh/cm$^2$ of reversible capacity at 50μA/cm$^2$ and maintains 65% of this capacity even when operated at 1 mA/cm$^2$. The ALD LPZ is therefore conductive enough that the batteries are rate-limited by solid state diffusion in the electrodes (indicated by the sharp rollover of the potential-capacity curve) well before any Ohmic drop in the electrolyte induces a voltage cut-off. These batteries have reasonable cycling stability, shown out to 150 cycles at a current of 300 $\mu A/cm^2$ in Figure 7c. The steady state reversible capacity is only ~50% of the theoretical capacity of the LCO (69 $\mu A/cm^2$ $\mu m$) after the first few cycles, which is due to irreversible processes associated with the Si anode (also observed in similar cells made with conventional RF-sputtered LiPON, Figure S5). The Coulombic efficiency stabilizes at approximately 99.8%, consistent with a very slow decline in capacity observed after the first few cycles. Li-ion SSBs are much more sensitive to Li loss than typical Li-anode thin film batteries, which utilize an effectively infinite Li reservoir. SEM characterization (Figure S6) suggests that some of the capacity loss is due to Li loss via diffusion through the copper current collector and reactions with environmental contaminants, including trace $H_2O$ and $O_2$ in the glovebox.

One of the primary advantages of ALD over other thin film growth methods is the ability to form high-quality, pinhole-free layers at very small thicknesses. RF sputtering requires careful optimization to avoid the formation of device-shorting defects, and most thin film SSBs utilize ~1 μm of electrolyte as a result. Nevertheless, there are a few examples of pinhole-free LiPON films grown on smooth metal surfaces down to a thickness of 12nm using sputtering or ion beam deposition, and one instance of 100nm LiPON working in a full battery.[41,59,60] Figure 7d shows a charge-retention experiment for a 350nm LCO/400 cycle LPZ-300/40nm Si full cell. 400 cycles of the LPZ-300 ALD process nominally produces a 36nm thick film based on a 0.9 Å/cyc growth rate, and FIB/SEM characterization of the LPZ/Si layer (inset) suggests that the LPZ thickness is between 30 and 35 nm, which is comparable to the peak-to-trough height of the rough, columnar LCO surface. After charging to 4V, the open circuit potential of these cells initially quickly drops due to the relaxation of internal concentration gradients but remains above 3.6V for the duration of the 12 hour test, indicating the tolerance of ultrathin ALD LPZ to progressive electrochemical decomposition at realistic solid state battery potentials as well as a tolerance to dielectric breakdown at field strengths of over 1 MV/cm. We note that a 35nm film of pristine LPZ-300 integrated into a battery would have a nominal ionic resistance of only 5.4 $\Omega \cdot cm^2$ at 35C. The above results demonstrate that ALD LPZ can act in a similar role as conventional sputtered LiPON at significantly smaller dimensions (and eventually in 3D topography, considering the demonstrated conformality) as a result of the excellent film quality.

## A Proof-of-Concept Flexible Solid State Battery

A promising area of applications for thin film solid state batteries is in flexible and wearable electronics, where the intrinsic safety and tailorable form factor of solid state storage is highly desirable. The relatively low temperature of the LiOtBu-DEPA ALD process, which produces high quality electrolytes at reasonable growth rates at 250C, allows the use of flexible metallized polyimide as a substrate (Figure 8a). As LCO requires a 700C annealing step to form the high performance crystalline phase, we replaced it with crystalline ALD $V_2O_5$, grown at 170C using a vanadium triisopropoxide and ozone process, which was subsequently electrochemically lithiated in a $LiClO_4$/propylene carbonate electrolyte to form $LiV_2O_5$.[61] Afterwards, 1500 cycles of LPZ-250 and 40 nm of evaporated Si completed a set of flexible solid state batteries, shown schematically in Figure 8b. Cyclic voltammetry of a $LiV_2O_5$/LPZ-250/Si full cell between a cell potential of 1 and 3.8 V revealed the oxidation/reduction peak doublet characteristic of $LiV_2O_5$, although the response is again broadened due to the use of Si as both the counter and reference electrode (Figure 8c). The pair of peaks labelled (1) is associated with the ε-δ transition ($LiV_2O_5 \leftrightarrow Li_{0.5}V_2O_5$) and the pair labelled (2) indicate the ε-α transition ($Li_{0.5}V_2O_5 \leftrightarrow V_2O_5$).[62] To demonstrate the batteries' tolerance to moderate bending, we galvanostatically cycled one sample 11 times at a current density of 50 μA/cm2, then removed it and flexed/unflexed it around a 1cm bending radius ten times before resuming cycling. As shown in Figure 8d, the battery remained electrically insulating and maintained a steady capacity of approximately 1.6 $\mu Ah/cm^2$. Similar to the batteries using LCO, the $LiV_2O_5$-based devices also experience a significant first-cycle capacity loss to the Si.

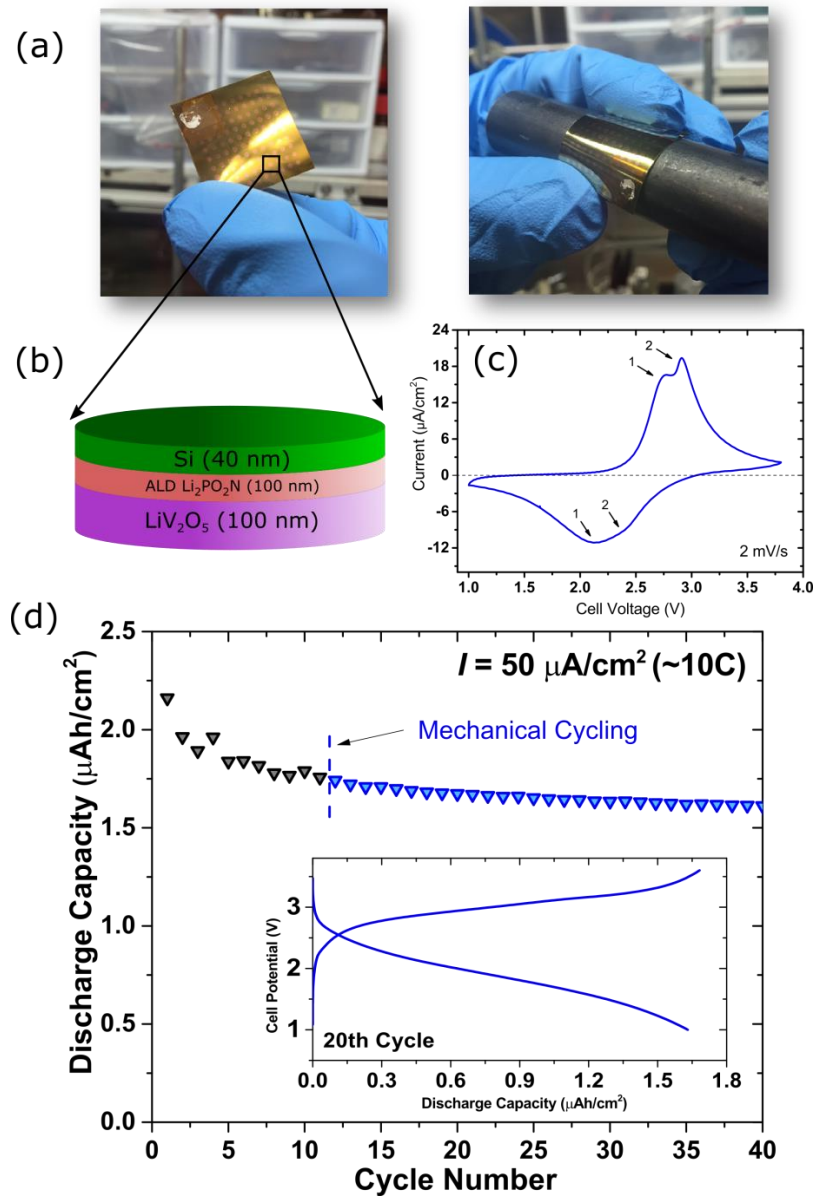

**Figure 8:** Proof-of-concept flexible solid state battery demonstrating the compatibility of ALD-LPZ with a polymer substrate. (a)Photographs of a sample, both unflexed and flexed, with an array of solid state cells. The small dots are individually defined top electrodes consisting of evaporated Si and Cu. (b) A schematic of the battery chemistry utilized (c) Cyclic voltammetry of a $LiV_2O_5$/LPZ-250/Si couple from a representative sample (in this case, grown on a Si substrate) showing two pairs of peaks associated with the lithiation/delithiation of the $V_2O_5$ cathode. The labelled arrows indicate a pair of cathodic/anodic peaks indicative of phase transitions in the $LiV_2O_5$ cathode. (d) Cycling stability of the battery stack shown in (b) on a polyimide substrate. The sample was removed after the 11$^{th}$ cycle and flexed 10 times around a 2cm bending radius rod before being replaced and cycled further. The inset shows a typical charge/discharge curve. The theoretical capacity of the battery, which uses 90nm LiV2O5 as a cathode, is approx.. 4.4 $\mu Ah/cm^2$, though the battery only shows 36% of this value as reversible capacity due to losses to the Si anode.

While the absolute capacity of these devices is too small to be practically useful, we believe the demonstration of the compatibility of the LPZ ALD process with both (1) a flexible polymer substrate and (2) an ALD-grown cathode provides a partial path towards the development of conformally grown solid state batteries in flexible systems of arbitrary geometries.

## Discussion and Conclusions

This work demonstrates two fundamental advances. First, a new thermal process chemistry for a LiPON-like ALD SSE is developed, leading to a chemical structure which seems best described as a lithium-conducting polyphosphazene. Second, we demonstrate for the first time that an ALD-grown solid electrolyte can work effectively in full batteries, opening the door to nonplanar SSB architectures.

The ALD reaction between LiO$^t$Bu and DEPA is self-limiting, and results in highly conformal solid electrolytes with reasonable growth rates between 250 and 300C, a temperature range low enough to enable the use of certain polymer substrates for flexible devices. In addition, the process utilizes only two precursors to produce a 4-element film, which keeps production times low. The LPZ films produced by the LiO$^t$Bu-DEPA reaction exhibit many attractive properties for use as thin film solid state electrolytes, including excellent electrochemical stability, reasonable ionic conductivity, and compatibility with two common cathode materials. The reaction is chemically surprising for a number of reasons, including the resulting stoichiometry of the LPZ films when compared to its precursors, and because it appears to differ significantly from a separately reported process which differs only in the Li ligand (a *tert*-butoxide group instead of a hexamethyldisilazide group). The chemical mechanism of the growth process and the origin of this difference deserve further investigations using *in-operando* chemistry-sensitive techniques, such as in-line mass spectrometry of the reaction byproducts or Fourier-transform infrared spectroscopy.

By integrating our the ALD LPZ into realistic solid state batteries, we have shown for the first time that an ALD-grown film can act as a drop-in replacement for sputtered LiPON. This is a necessary step in demonstrating the viability of 3D solid state batteries, which universally require defect-free conformal and stable SSEs. In some respects, the ALD-grown electrolyte is superior to sputtered films in that it can provide electrical isolation at thicknesses smaller than those previously reported in the literature by a factor of ~3, increasing the overall energy density. We have also shown that while the ALD LPZ is air reactive through the formation of a $Li_2CO_3$ decomposition layer, this does not seriously impede its use as an electrolyte. The combination of reasonable ionic conductivity and reliable film closure at small thicknesses, even on rough substrates, allows reliable fabrication of electronically insulating < 20 $\Omega \cdot cm^2$ solid lithium electrolytes via ALD.

## Experimental Methods

### ALD Growth

All ALD processes were performed in a custom Cambridge Nanotech Fiji F100 ALD reactor directly coupled to an ultrahigh vacuum cluster tool. A schematic of the cluster tool is shown in Figure S1. All processes used UHP (99.999%) Ar as the process gas, typically achieving a background pressure of ~200 mTorr during deposition. Depositions of $Li_2PO_2N$ utilized $LiOC(CH_3)_3$ referred to as lithium *tert*-butoxide or LiO$^t$Bu, (Sigma) and $H_2NPO(OC_2H_5)_2$, referred to as diethyl phosphoramidate or DEPA (Sigma). Both materials are solids at room temperature. LiO$^t$Bu was stored in a stainless steel bubbler, heated to 140C, and delivered to the reactor by co-flowing 15 sccm of Ar. LiOtBu decomposes at approximately 320C.[32] The LiO$^t$Bu delivery lines include VCR particle filters to prevent fine particles of precursor from reaching the chamber, which was an issue for early devices. DEPA did not require bubbling and was stored in a conventional stainless steel ALD cylinder heated to 115C. Unless otherwise specified, the pulse and purge times used for depositions in this work were 20s-LiO$^t$Bu, 20s-purge, 2s-DEPA, 20s-purge. Some samples utilized an "exposure" process in which a butterfly valve shut off all active pumping to the ALD chamber during precursor exposure to allow for better conformality. The timing of this process was 10s-LiO$^t$Bu (10s exposure), 30s-purge, 2s-DEPA (10s exposure), 20s-purge, and exhibited very similar growth characteristics to the conventional process.

**In-situ Ellipsometry**

In-situ ellipsometry was taken using a J.A. Woollam M-2000 spectroscopic ellipsometer. The source and collector heads were mounted to quartz windows on the ALD reactor at a fixed angle. All optical models were applied to a spectral range of λ = 300-1000 nm. The deposited films were optically modelled as transparent insulators using the Cauchy approximation $n(\lambda) = A + B\lambda^{-2} + C\lambda^{-4}$ where $n$ in the index of refraction, $\lambda$ is the wavelength of light, and A, B, and C are fitting constants.[63] Consistent with previous reports for LiPON, the SE data for the films were well fitted with $A \approx 1.7$, and B and C $\approx 0$, indicative of a nearly constant index of refraction over the measured bandwidth.[26,64] We also assume $k(\lambda) = 0$, where $k$ is the absorption coefficient. The optical model was externally verified via comparison with x-ray reflectivity (XRR) measurements and SEM/FIB cross sections of various reference samples, and all thickness measurements agreed to within 5%.

**XPS Analysis**

Samples were immediately transferred under ultrahigh vacuum from the ALD chamber to a customized Kratos Ultra DLD x-ray photoelectron spectrometer with a base pressure of 2 × 10$^{-9}$ torr. This preserves the surface chemistry of air-reactive Li compounds and allows for accurate stoichiometric quantification. All XPS data was collected using monochromatic Al Kα radiation (1486.7 eV) at a total power of 144W. The analysis spot size was approximately 0.2 mm$^2$. Survey and high-resolution spectra were collected using 160 eV and 20 eV pass energies, respectively. Samples were not observed to change over time in the vacuum environment. CasaXPS was utilized for peak fitting (using 50/50 Gaussian/Lorentzian pseudo-Voigt functions) and data analysis. High resolution peak area ratios were used for elemental quantification, using tabulated Kratos relative sensitivity factors (Scofield cross sections corrected for the instrument transmission function and source-analyzer angle). All spectra were calibrated to the C 1s hydrocarbon peak at 284.8 eV, though this assignment has associated uncertainty as the hydrocarbons in this case are embedded fragments and not adsorbed species. Depth profiles were performed using a Kratos Gas Cluster Ion Source (GCIS) on a Kratos AXIS Supra spectrometer for sample sputtering using $Ar_n^+$ cluster ions, which proved superior to monoatomic Ar sputtering sources for best preserving the stoichiometry of LPZ films.

**Microscopy and Characterization**

Scanning electron microscopy (SEM) and focused ion beam (FIB) work was performed using a Tescan GAIA dual SEM/FIB system, which includes an attached TOF-SIMS detector used for the detection of Li during depth profiling with the Ga$^+$ ion beam. Transmission electron microscopy work was performed using a JEM 2100 FEG TEM. The ALD LPZ was found to be highly sensitive to beam damage in the TEM and exposures were kept as short as possible. All imaged battery samples, including the TEM lamella, were exposed to air for several minutes during transfer from system to system. Tapping-mode AFM was performed using a NT-MDT NTEGRA Specta and XRD was checked using 1000 cycle films deposited on Au using a Bruker C2 Discover.

**Device Fabrication**

Multiple architectures were utilized in this study. *In-situ* ALD growth was characterized on RCA-cleaned Si test wafers. Devices were fabricated on diced thermally oxidized Si wafers. Metal depositions for current collectors and MIM electrodes (including Pt and Au) were performed using electron-beam physical vapor deposition (EBPVD), utilizing a 5nm Ti or Cr adhesion layer for the bottom electrode. $LiCoO_2$ electrodes were fabricated by RF sputter deposition of a LiCo target under flow of Ar and $O_2$ in a 3:1 ratio, and were annealed at 700C. $LiV_2O_5$ electrodes were fabricated by first growing $V_2O_5$ in a Beneq TFS 500 ALD reactor at 170C using vanadium triisopropoxide (VTOP) and $O_3$ and subsequently electrochemically lithiating the films to a potential of 2.8V vs. $Li/Li^+$ in a 0.5M $LiClO_4$/propylene carbonate electrolyte using a Li metal counterelectrode. Excess electrolyte was rinsed off using ethanol, and the composition was verified using XPS. To form an electrical contact, one corner of each device was masked during both cathode deposition and $Li_2PO_2N$ deposition by physically clamping a piece of a silicon wafer to the surface. Top electrodes were deposited through a stainless steel shadow mask which defined a grid of 1mm diameter circular pads, which determined the active device area. Li top electrodes were deposited using thermal evaporation of Li metal pieces (Sigma) in a vacuum chamber directly connected to a Ar-filled glove box. Si/Cu top contacts were deposited in one process without breaking vacuum using EBPVD at a pressure of $3 \times 10^{-6}$ torr, but these samples were air exposed for several minutes after electrolyte growth for transport to the deposition tool. Flexible devices were fabricated on cut pieces of metallized polyimide sheet, using evaporated Au with a Cr adhesion layer for a bottom electrode.

**Electrochemical Characterization**

Fabricated devices were tested in an Ar-filled glovebox with <0.1 ppm $H_2O$ and $O_2$ using a homebuilt microprobe setup. The sample is clipped to a stage with an integrated PID temperature control unit and a metal clip is used to contact the bottom electrode. The top electrode is contacted via an Au-coated needle probe mounted to a micromanipulator. Both electrodes are then connected to a Biologic VSP potentiostat with an electrochemical impedance spectroscopy channel using a coaxial cables and BNC feedthrough. Unless otherwise specified, measurements were taken at ambient temperature (typically 27C). PEIS measurements were taken between 1MHz and 0.1 Hz with an excitation amplitude of 50mV.

ASSOCIATED CONTENT

**Supporting Information**. Schematic of experimental setup, detailed PEIS fitting parameters, further in-situ SE data, XPS depth profiles of air-exposed films, a comparison to similar devices using sputtered LiPON, and SEM characterization of cycled batteries.

AUTHOR INFORMATION

**Corresponding Author**

*Alexander J. Pearse: ajpearse@umd.edu

*Keith E. Gregorcyzck: kgregorc@umd.edu

**Author Contributions**

A.J.P., G.R., K.G., and K.E.G. developed concepts and planned research. A.J.P. and K.E.G. developed and characterized the ALD process. A.J.P., K.E.G., and T.E.S. fabricated and characterized the solid state cells and half-cells. The $LiCoO_2$ thin film cathodes were developed and fabricated by E.J.F., F.E., and A.A.T. C.L.

assisted with characterization. A.J.P. wrote the manuscript. A.J.P. and A.C.K. initiated ALD solid electrolyte research. All authors discussed and approved the final version of the manuscript.

ACKNOWLEDGMENT

All aspects of this work were initiated and supported by Nanostructures for Electrical Energy Storage (NEES), an Energy Frontier Research Center (EFRC) funded by the U.S. Department of Energy, Office of Science, Office of Basic Energy Sciences under award number DESC0001160. A flexible version of a solid state battery was included as an additional prototype demonstration, partially supported by Independent Research and Development Funding from the Research & Exploratory Development Department (REDD) of the Johns Hopkins University Applied Physics Laboratory (JHU/APL). We appreciate the facilities and support within the Maryland NanoCenter, including its Fablab for device fabrication and its AIMLab for microscopy and FIB. We also acknowledge Jonathan Counsell of Kratos Analytical for assistance with XPS depth profiling using a cluster ion source, as well Mr. John Fite from REDD at JHU/APL for his experimental assistance. Sandia National Laboratories is a multi-mission laboratory managed and operated by Sandia Corporation, a wholly owned subsidiary of Lockheed Martin Corporation, for the U.S. Department of Energy's National Nuclear Security Administration under contract DE-AC04-94AL85000.

# Supplementary Information

**Figure S1:** Diagram of the Integrated System

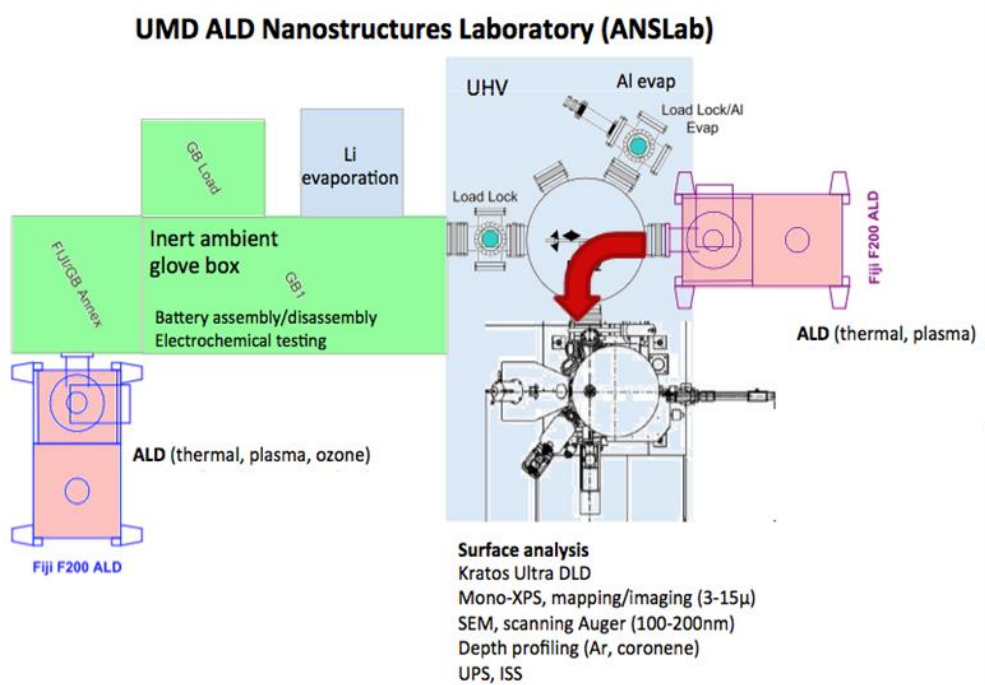

**Figure S1:** A schematic of the integrated vacuum/deposition system used for experiments described in the main text. ALD-grown samples can be transferred directly from the ALD chamber to the XPS in UHV via the red arrow.

**Table S1:** Detailed Fitting Parameters for PEIS taken at 35C

| Sample ID | LPZ Thickness (nm) | $R_b$ (Ω) | $R_r$ (Ω) | $CPE_b$ Q | $CPE_b$ n | $CPE_r$ Q | $CPE_r$ n | $\chi^2$ |
|---|---|---|---|---|---|---|---|---|
| LPZ-300 (Pt/Pt) | 80 | 1085 | 5386 | $9.02 \times 10^{-9}$ | 0.899 | $9.16 \times 10^{-8}$ | 0.795 | $2.3 \times 10^{-4}$ |
| LPZ-250 (Pt/Pt) | 70 | 1448 | 5049 | $7.93 \times 10^{-9}$ | 0.903 | $1.32 \times 10^{-7}$ | 0.783 | $2.4 \times 10^{-4}$ |
| LPZ-300 (Pt/Li) | 90 | 1749 | — | $8.91 \times 10^{-9}$ | 0.864 | — | — | $1.3 \times 10^{-4}$ |

Table S1 lists the parameters used to model the ionic conductivity of ALD LPZ films at 35C, which includes a component associated with the film bulk ($R_b$ and $CPE_b$) and the $Li_2CO_3$ reaction layer in the case of air exposed films ($R_r$ and $CPE_r$). The constant phase element impedance is a generalized capacitance and is described by the equation

$$Z = \frac{1}{Q\omega^n} e^{-i\pi n/2}$$

and the associated values for $Q$ and $n$ are listed in the table. $\chi 2$ describes the goodness of fit for each model.

**Figure S2:** In-Situ SE of DEPA Pulses Alone at 300C

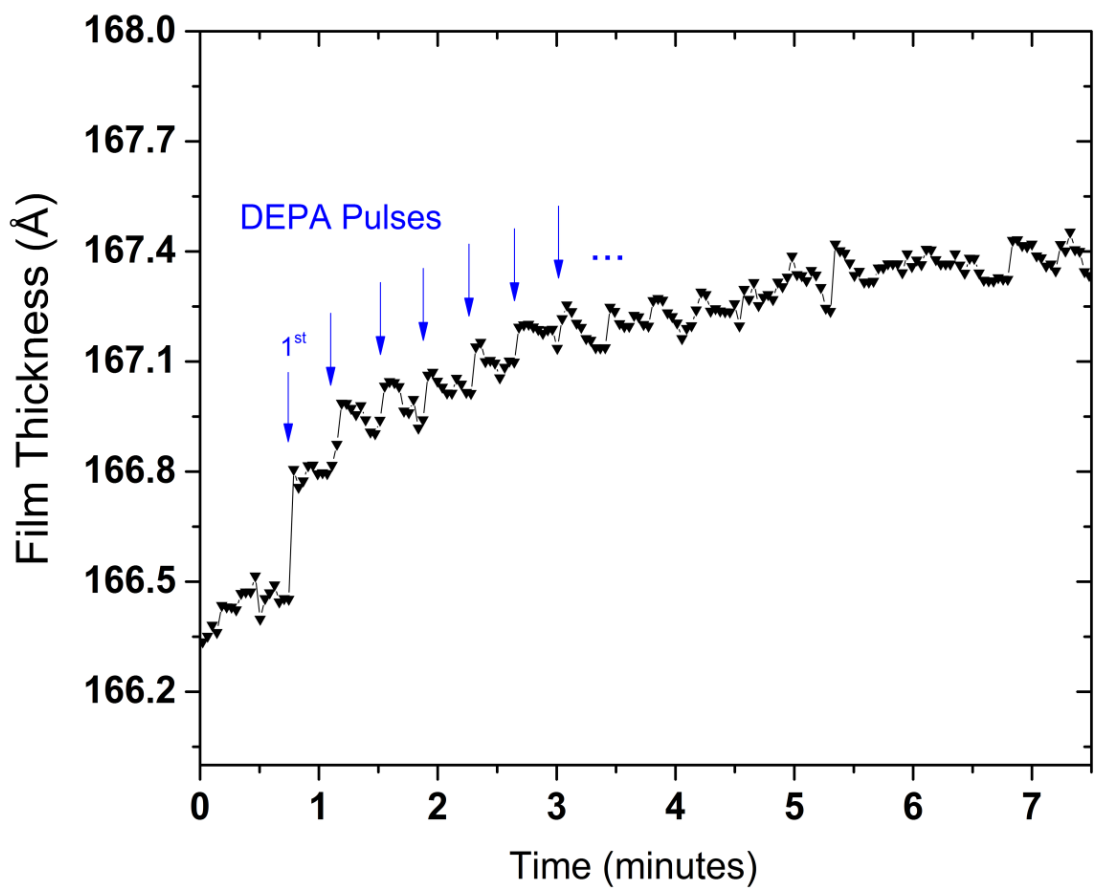

**Figure S2:** In-situ ellipsometry of film growth while pulsing DEPA only (no LiO$^t$Bu pulses) at 300C on a pre-grown LPZ-300 surface 16.6 nm in thickness. The blue arrows mark the locations in time of 2s DEPA pulses (separated by 30 seconds and which continue during the entire plotted time period). The first pulse results in the strongest change measured by SE, and subsequent pulses contribute smaller and smaller differential changes in film thickness before saturating at a total change of ~ 1 angstrom. This indicates that DEPA alone does not form a film.

**Figure S3:** Gas Cluster Source XPS Depth Profile of Air-Exposed ALD LPZ

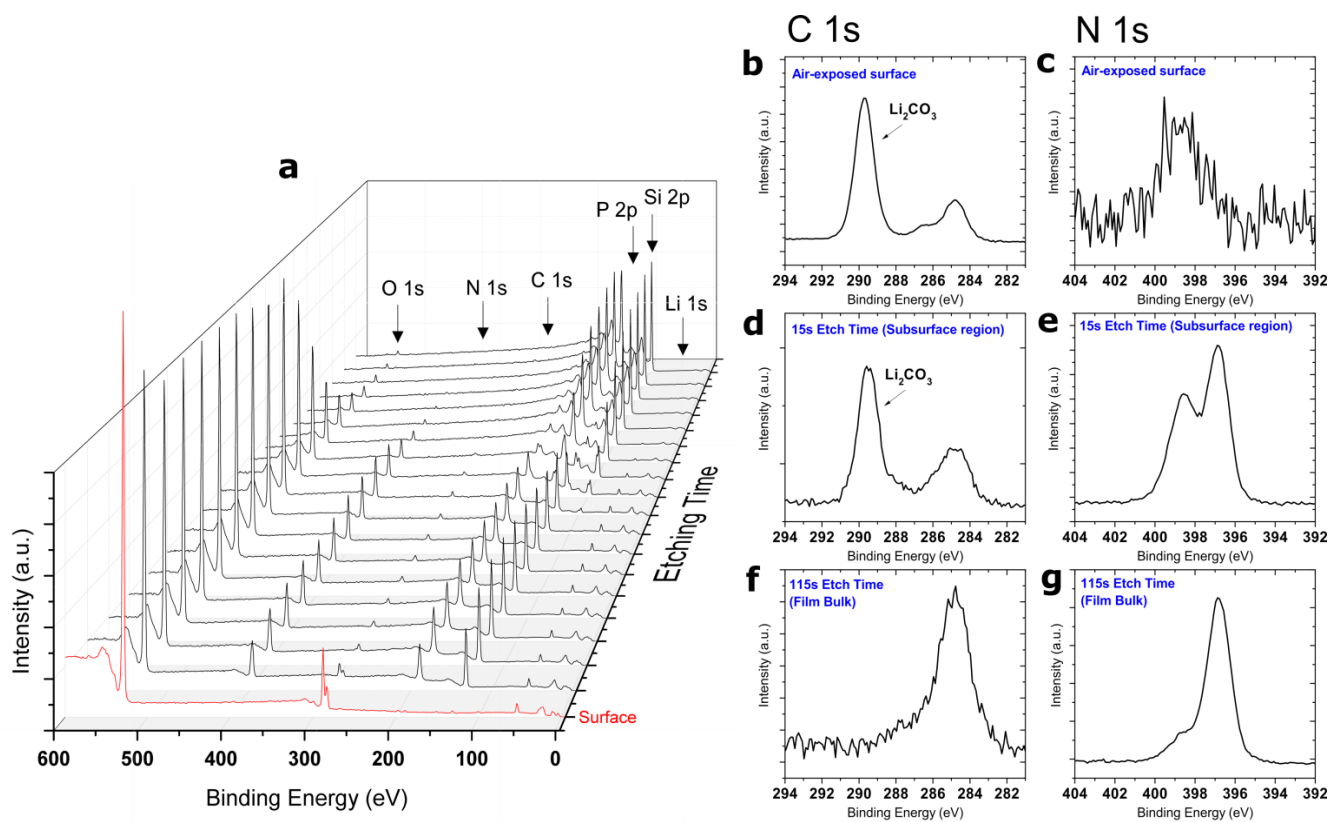

**Figure S3:** Data from an XPS depth profile of an air-exposed ALD LPZ-300 film grown on a Si substrate (total thickness ~50nm). (a) Survey spectra plotted as a function of etching time (b-g) High resolution spectra of the C 1s and N 1s core levels taken at the surface (b,c) after one etch cycle (d,e) and from the bulk of the film (f,g). High resolution spectra are calibrated to the hydrocarbon component at 284.8 eV.

Figure S2 shows the presence of a decomposed/reacted surface layer formed in air-exposed ALD LPZ films. The C 1s core level spectra taken from the surface (Figure S3b) and just below the surface (Figure S3d) clearly show the presence of a carbon species at 289.7 eV, which is characteristic of lithium carbonate.[1] The N chemistry of the film is also affected near the surface, with increasing N content associated with the N ε component identified in Figure 2 in the main text. After several etching cycles, the chemistry again resembles that measured for pristine films (Figure S3f,g). The thickness of the reaction layer can be estimated to be on the order of a few nm.

We found that normal monoatomic ion sputtering sources significantly degraded the film. As a result, the XPS/GCIS analyses were carried out with a Kratos AXIS Supra spectrometer equipped with a Gas Cluster Ion Source (GCIS) for sample sputtering using $Ar_n^+$ cluster ions or monatomic $Ar^+$. Clusters are created in the GCIS via the supersonic expansion of high pressure Ar gas through a de Laval nozzle into a medium vacuum region. Nascent clusters are transmitted through a differentially pumped region into an electron impact ionisation source. Following ionisation, $Ar_n^+$ cluster ions are extracted through the length of a Wien filter to eliminate all small ions and limit the transmitted cluster size distribution spread around a chosen median value. The $Ar_n^+$ cluster ions are then deflected through 2° to eliminate neutral and metastable species from the beam before simultaneous focusing and rastering across the sample. Alternately, Ar gas can be directly introduced into the electron impact ionisation source to allow production of monoatomic $Ar^+$ ions. The GCIS can deliver $Ar_n^+$ cluster ions at energies from 0.25–20 keV and cluster sizes n = 150–5000.

**Figure S4:** TEM Damage of ALD LPZ

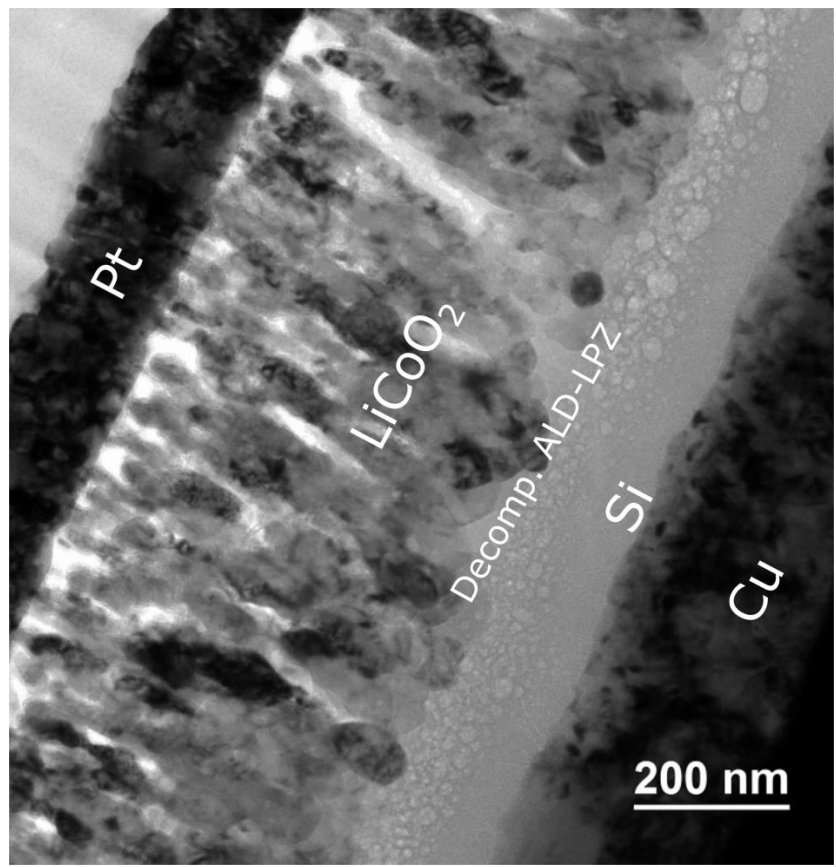

**Figure S4:** TEM image of a cross-sectioned LCO/LPZ-300/Si battery after extended exposure to the electron beam. The LPZ layer decomposes and forms bubbles, likely through gas evolution. This allows for easy visual differentiation of the LPZ and Si layers.

**Figure S5:** First cycle capacity loss in LCO/LiPON/Si batteries using RF-sputtered LiPON

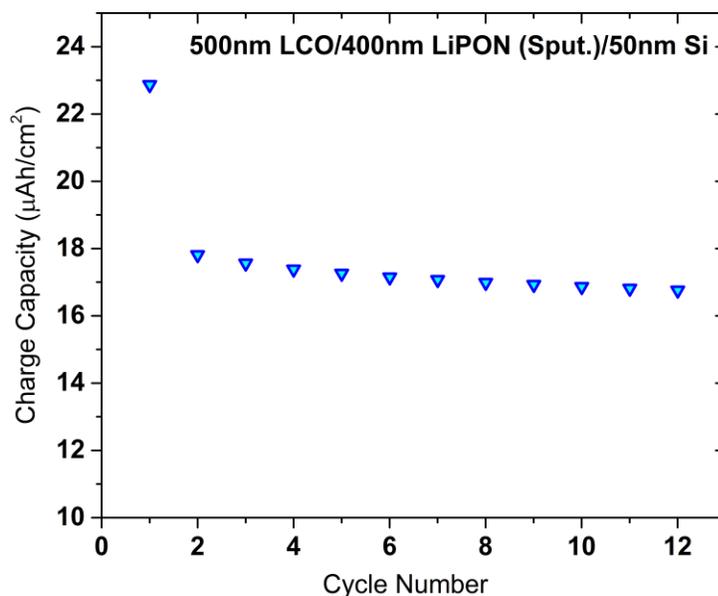

**Figure S5:** Charge capacity at a C/10 rate of a SSB made with 500nm LCO, 400nm RF-sputtered LiPON (made using previously published procedures), and 50nm Si.[2] The battery was cycled between 3 and 4.2V. These devices show a similar first cycle capacity loss and stable capacity to batteries made using ALD LPZ, which indicates the majority of the irreversible capacity is likely associated with the Si anode and not the electrolyte.

**Figure S6:** SEM of Surface Reaction Layer in Cycled Batteries

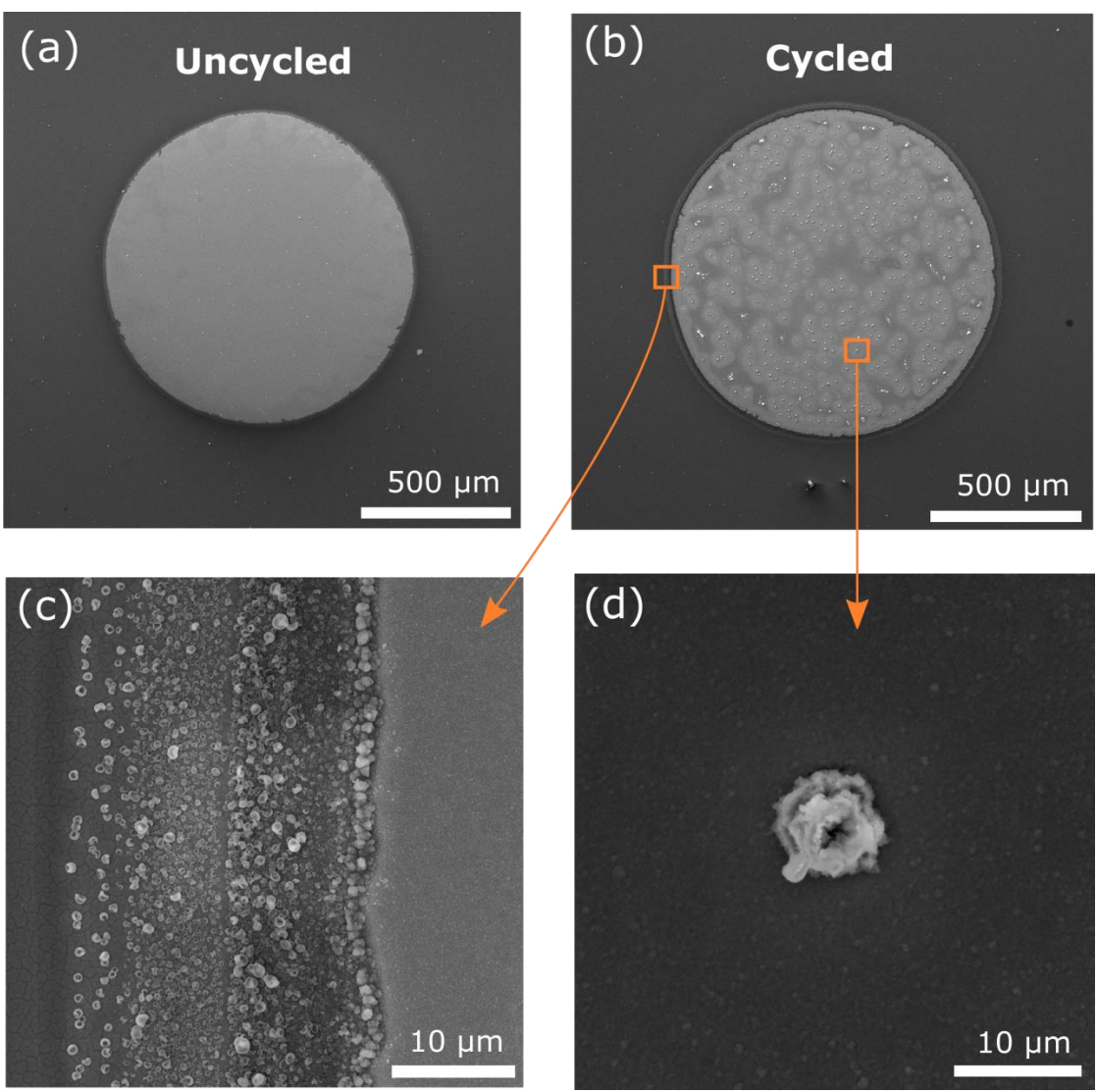

**Figure S6:** SEM images of (a) an as made, uncycled battery and (b) a cycled battery. The battery chemistry is identical to the devices described in Figures 6 and 7 of the main text. Cycled devices evolve debris both on the surface (d) and near the edges (c) of the Cu current collector, which we believe to be lithium compounds formed from the reaction of mobile lithium inserted into the Si anode and environmental contaminants, including trace $H_2O$ and $O_2$ in the glovebox. This may be one source of capacity loss over time, and future devices will be encapsulated to help prevent this effect.

**Figure S7:** XRD of LPZ-300

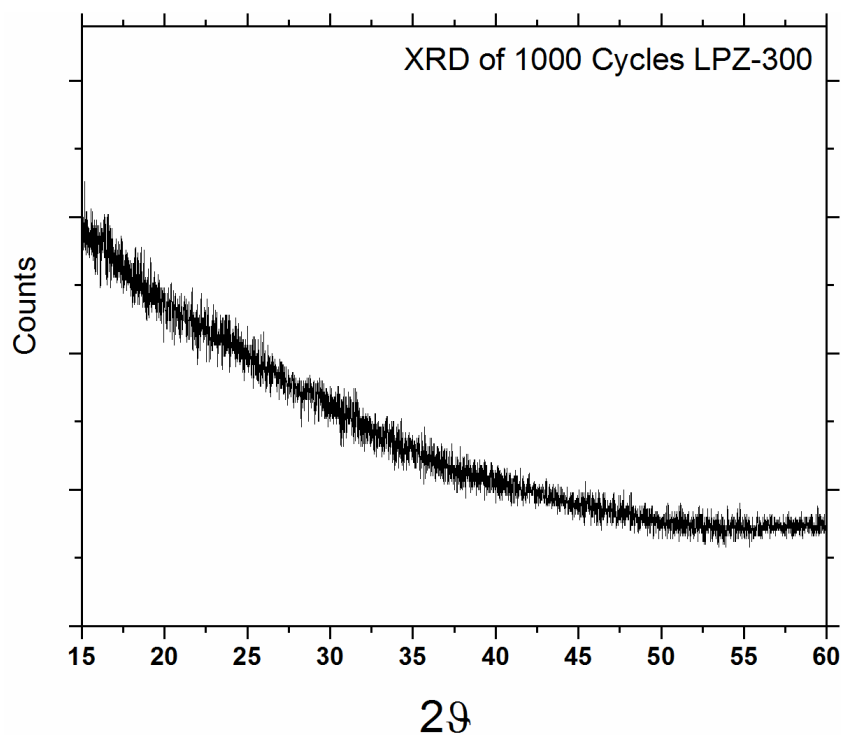

**Figure S7:** XRD of a LPZ-300 sample showing the amorphous nature of the films. This corresponds well to the lack of structure identifiable in TEM cross section studies.

## Supplemental References